# Oxygen adsorption effect on magnetic properties of graphite

D. W. Boukhvalov[1,2], S. Moehlecke[3], R. R. da Silva[3], and Y. Kopelevich[3]

[1]*School of Computational Sciences Korea Institute for Advanced Study Hoegiro 87, Dongdaemun-Gu, Seoul, 130-722, Korean Republic*

[2]*Computational Materials Science Center, National Institute for Materials Science, 1-2-1 Sengen, Tsukuba, Ibaraki 305-0047, Japan*

[3]*Instituto de Física "Gleb Wataghin", Universidade Estadual de Campinas, Unicamp 13083-859, Campinas, São Paulo, Brasil*

Both experimental and theoretical studies of the magnetic properties of micrographite and nanographite indicate a crucial role of the *partial* oxidation of graphitic zigzag edges in ferromagnetism. In contrast to total and partial hydrogenation, the oxidation of half of the carbon atoms on the graphite edges transforms the antiferromagnetic exchange interaction between graphite planes and over graphite ribbons to the ferromagnetic interaction. The stability of the ferromagnetism is discussed.

E-mail: danil.@kias.re.kr

E-mail: kopel@ifi.unicamp.br

The occurrence of ferromagnetic (FM)-type behavior above room temperature in metal-free carbon-based (CB) structures has been previously reported by various research groups [1]. In most cases, the measured magnetic moment has been rather small, so care must always be taken regarding the effects of possible traces of ferromagnetic impurities, such as Fe, Co, and Ni [2]. Nevertheless, there exists convincing experimental evidence supporting the intrinsic character of ferromagnetism in CB materials [1].

In particular, Murakami and Suematsu [3] produced ferromagnetic order in $C_{60}$ fullerene crystals by irradiating them with light from a xenon lamp in the presence of oxygen. They showed that the diamagnetism characteristic of fullerenes had been overwhelmed by paramagnetic or ferromagnetic responses after irradiation of the sample in an oxygen atmosphere. The decrease of the ferromagnetic moment after annealing and its increase after leaving the sample in air for several months argue against the contributions of Fe impurities. The authors estimate a magnetic moment of 0.1 $\mu B$ per $C_{60}$ molecule. The temperature dependence of the saturation magnetization revealed an extraordinarily high Curie temperature, $T_c = 800$ K. Makarova *et al.* [4] reported similar effects for laser- and electron-beam-illuminated $C_{60}$ films obtained either in air or in an oxygen-rich atmosphere. Importantly, x-ray structural analysis performed on those $C_{60}$ samples revealed the coexistence of the $C_{60}$ phase with clusters of graphene layers [4]. This fact is notable because of the occurrence of ferromagnetism in bulk graphite [5-9].

Bearing in mind that in all cases, magnetism in CB materials is associated with graphene layers, it is instructive to explore the role of zigzag edges [8-14]. In fact, the results of recent experiments [8, 15] speak for the key role of graphite zigzag edges in magnetism. In line with these experiments, zigzag half-metallic antiferromagnetism [11, 12], energy-gap opening [13], and spiral-like magnetism [14] have been theoretically predicted. It is worth noting that, in contrast to unstable graphene zigzag edges [15, 16], the same edges in graphite are stable at room temperature [8, 17]. Theoretical studies also suggest that graphene-edge passivation affects magnetism [18, 19]. Both theoretical [19] and experimental [20, 21] work indicates the enormous oxygen affinity of zigzag graphene edges, and complete oxidation of these edges is expected to lead to a non-magnetic state due to the saturation of dangling bonds [19].

In contrast to graphene, zigzag edges in graphite do not reconstruct under non-oxidative cut. The production of graphite powder and/or nanographites in an oxygen environment is the method used to explore the role of oxygen in the edge magnetism of graphite.

In this work, we report the results of density functional theory (DFT) calculations for the effect of oxygen on the zigzag edge magnetism of graphene multilayers. We also describe our experimental studies on the effect of oxygen and other adsorbed gases on the magnetic properties of a graphite powder.

Density functional theory (DFT) calculations were carried out with the pseudopotential code SIESTA [22]. All calculations were performed using the local density approximation (LDA) [23], which is more suitable for the description of graphite [24] and was used for our previous modeling of graphite functionalization [25]. All calculations were carried out for an energy mesh cut-off of 360 Ry and k-point mesh $4 \times 8 \times 4$ in the Mokhorst-Park scheme [26]. During optimization, the electronic ground state was found self-consistently using norm-conserving pseudo-potentials for cores and a double-$\zeta$ plus polarization basis of localized orbitals for carbon and iron. Optimization of the forces and of total energies was performed with accuracies of 0.04 eV/Å and 1 meV, respectively. When creating images of the density of states, a smearing of 0.2 eV was used.

To model realistic graphite, the nanoribbon was multiplied along the z-axis, using periodic boundary conditions and taking into account AB (Bernal) stacking (Fig. 1a). The total number of carbon atoms in the studied nanographite supercell is 128. To explore the specific role of oxygen, we modeled total (Fig. 1c) and different types of partial (Fig. 1d-g) hydrogenation and oxidation.

Exchange energy is defined by the standard formula $J = (E_{FM} - E_{AFM})/N$, where N is the number of zigzag edge atoms with magnetic moments in the studied supercell. The Curie temperature is calculated using the equation $T_C = JS(S+1)/3k_B$, where J is the sum of exchange energies in all three (x, y, and z) directions (see Fig. 1) of each magnetic atom on the zigzag edge, and S is the spin per zigzag edge step.

An activated graphite powder was prepared by cutting and grinding a graphite rod at T = 300 K in different atmospheres: Ar, He, $N_2$, $H_2$, $O_2$, and air. The graphite rod was from Carbon of America Ultra Carbon, sold by Alfa Aesar (stock No. 40766), AGKSP grade, (ultra "F") 99.9995 % purity or 5 ppm of total impurities and maximum of 1 ppm of impurities per element. The powder was produced by cutting and grinding the graphite rod on the edge and side area of a new, clean circular diamond saw blade. The cutting and grinding system was inside a plastic bag filled with the gas. A continuous gas stream was also forced to blow through a gas hose into the grinding area. From the x-ray diffraction spectrum, we estimated the graphite crystallite size to be $L_a \approx 1000$ Å and $L_c \approx 400$ Å. The powder particle size was found to be $150 \pm 70$ µm, using sieves.

Magnetization M(T,H) measurements were performed using a commercial SQUID (Quantum Design, MPMS5) magnetometer.

Figures 2 and 3 present our principal experimental observations. Fig. 2 (a) illustrates that while the virgin bulk sample is diamagnetic with a negligible magnetic hysteresis, the sample prepared under oxygen exposure possesses a pronounced ferromagnetic response at room temperature. Figure 2(b) demonstrates the intrinsic, i.e., related to the sample, ferromagnetism. This intrinsic ferromagnetism can also be seen from Fig. 3, which shows that no ferromagnetic response was detected for graphite powders prepared under He, $N_2$, or Ar gas environments, in contrast to the samples prepared in the air or in an oxygen atmosphere.

Figure 4 illustrates another important experimental fact, namely that the oxygen effect is reversible: the ferromagnetism vanishes with time after the sample is removed from the oxygen atmosphere. This observation provides unambiguous evidence that the ferromagnetism of graphite is essentially related to the presence of oxygen.

Figure 5 emphasizes that hydrogen does not induce ferromagnetism in graphite. This observation should be compared to the pronounced ferromagnetism observed after proton irradiation [27-28]. We attribute the difference in the results to the much weaker effect of adsorbed hydrogen on the electronic structure of zigzag graphitic edges (see below) than that produced by high-energy protons [29].

To explore the role of oxygen adsorption, cases of total (Fig. 1b), slight (Fig. 1c, d) and half oxidation (Fig. 1e, f) of graphite zigzag edges were analyzed. Optimized atomic structure and calculated energy differences per edge unit between ferromagnetic and antiferromagnetic configurations are reported in Fig. 1. In non-oxidized nanographite, much as in graphene [11, 12, 14], magnetic interactions between the nearest neighbors belonging to the same edge are ferromagnetic ($J_y$), but interactions across the ribbon ($J_x$) are weakly antiferromagnetic. The magnetic interactions between neighbors in different graphite layers are also antiferromagnetic. The electronic structure of the studied nanographite (Fig. 6) is similar to the half-metallic structure reported for the case of graphene nanoribbons [11]. The total oxidation of graphite edges (Fig. 1b), much as in the case of graphene [19], leads to the absence of magnetism due to the saturation of dangling bonds on the zigzag edges.

Partial oxidation of graphite edges combines the passivation of several dangling bonds with an increase in the number of the states on the Fermi level (metallization) of the samples, as noted before for the graphene case [20, 21]. As can be seen in Fig. 1c and d, a few oxygen atoms chemisorbed on the edges significantly reduce antiferromagnetic interactions over the ribbon and between graphite plates. Further oxidation of graphene edges enhances the metallicity of nanographites (Fig. 6) and turns all antiferromagnetic exchanges to ferromagnetic. Thus, due to partial oxidation of nanographite edges, one-dimensional ferromagnetism becomes two-dimensional (Fig. 1e) and three-dimensional (Fig. 1f). The Curie temperature estimated for the case reported in Fig. 1f is 333 K, and the Curie temperature estimated for the case with two-dimensional ferromagnetism (Fig. 1e) is 310 K. Thus, these theoretical results predict the room-temperature ferromagnetism in partially oxidized graphite powders, as observed experimentally.

To explore aging, i.e., time-dependent effects, calculations of the total energy of chemisorbed oxygen atoms for the structures given in Fig. 1c-f were performed using the formula $E_O = (E_{nr} - E_{pure})/N$, where $E_{nr}$ is the energy of the partially oxidized nanoribbon, $E_{pure}$ is the energy of non-oxidized nanoribbon (see Fig. 1a), and N is the number of oxygen atoms in the supercell. The lowest total energy of oxygen corresponds to the lowest oxygen content on the edges (Fig. 1c). Adsorption of the next oxygen atom (Fig. 1d) increases the total energy by about 59 meV/oxygen. The cases of half passivation (Fig. 1e, f) are significantly energetically unfavorable (475 and 387 meV/oxygen, respectively). Hence, placing the samples in an oxygen-free environment results in redistribution of adsorbed oxygen atoms from areas with a higher concentration to non-oxidized edges or in desorption of oxygen atoms, leading to the weakening or even vanishing of the ferromagnetism.

The studies reported here of ferromagnetism in micrographite and nanographite samples provide unambiguous evidence for the crucial role of oxygen adsorption in room-temperature ferromagnetism in graphite. We demonstrated experimentally that hydrogen, nitrogen, helium, and argon gases do not induce ferromagnetism in graphite. DFT calculations revealed significant changes in the electronic structure of graphite ribbons caused by *partial* oxidation, transforming antiferromagnetic interactions to ferromagnetic interactions in two- and three-dimensional graphites. At the same time, according to our calculations, partial hydrogenation of graphitic edges does not lead to significant changes in the electronic and magnetic structures of the nanographites. Experimental results revealed the aging phenomenon, i.e., vanishing of the ferromagnetism over time when the sample is removed from the oxygen atmosphere. Theoretical modeling explains this effect as a result of the reorganization of the oxygen atoms into more energetically favorable antiferromagnetic configurations. The reported results as a whole demonstrate that the electronic and magnetic properties of carbon nanosystems can be strongly affected by graphitic edge passivation, which may be useful for the production and theoretical modeling of magnetic graphitic nanoribbons.

**Acknowledgements** This work was supported by CNPq, FAPESP, and CAPES.

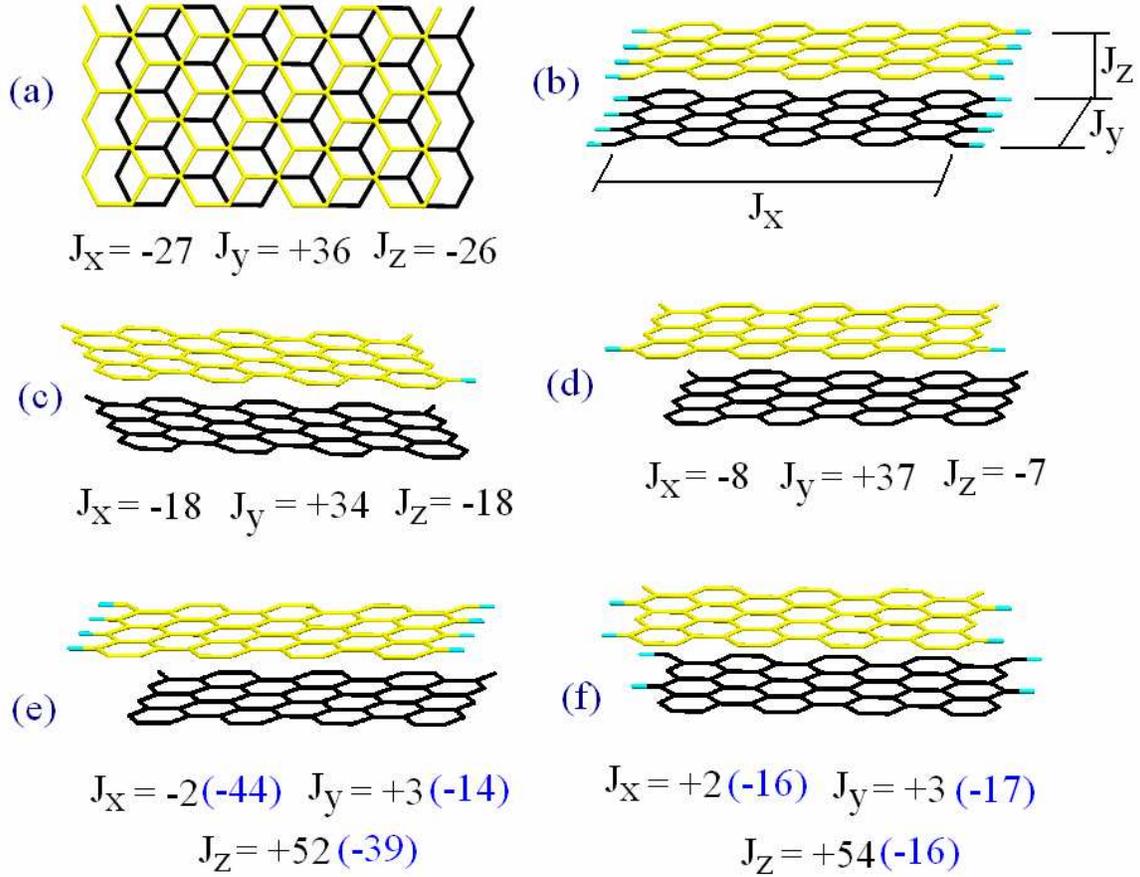

**Figure 1** (Color online) Optimized geometry structures of a nanographite sample with nonpassivated zigzag edges (a), with zigzag edges with total oxidation (b), and with different levels (1/16 (c), 1/8 (d), and 1/2 (e-f)) of oxidation. Carbon atoms from different graphite layers are shown in black and yellow (black and light gray), oxygen atoms in cyan (medium gray). All differences in energy between ferromagnetic and antiferromagnetic configurations are reported in meV. In parentheses, the same values are given for the case of hydrogenation of the edges.

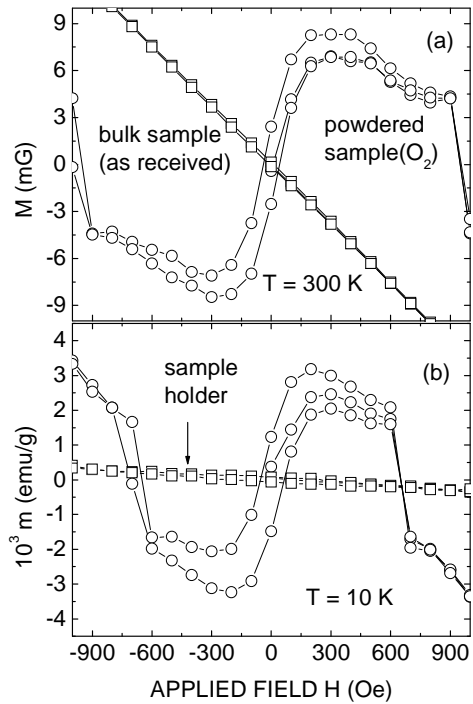

**Figure 2** (a) Magnetization M(H) measured at T = 300 K for the graphite samples: bulk (virgin) sample and powdered samples prepared under oxygen ($O_2$) atmosphere; (b) m(H) measured for the oxidized sample at T = 10 K, shown in comparison with the negligible contribution from the sample holder.

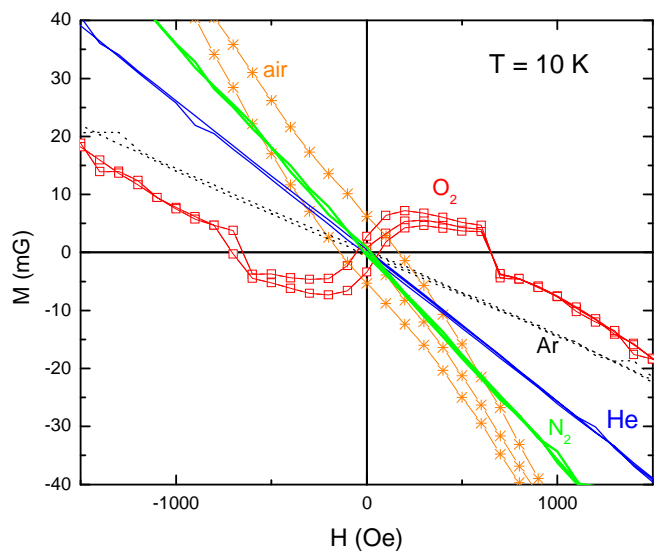

**Figure 3** (Color online) Magnetization M(H) measured at T = 10 K for the powdered graphite samples prepared in various gas environments (see text for details).

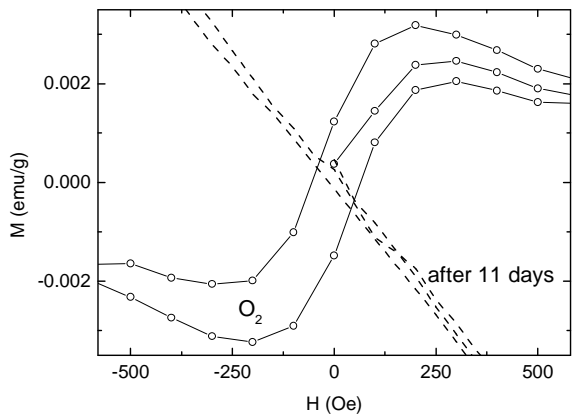

**Figure 4** M(H) for the graphite powder prepared in an oxygen ($O_2$) environment and measured at T = 10 K immediately after preparation and 11 days after the sample has been removed from the oxygen atmosphere.

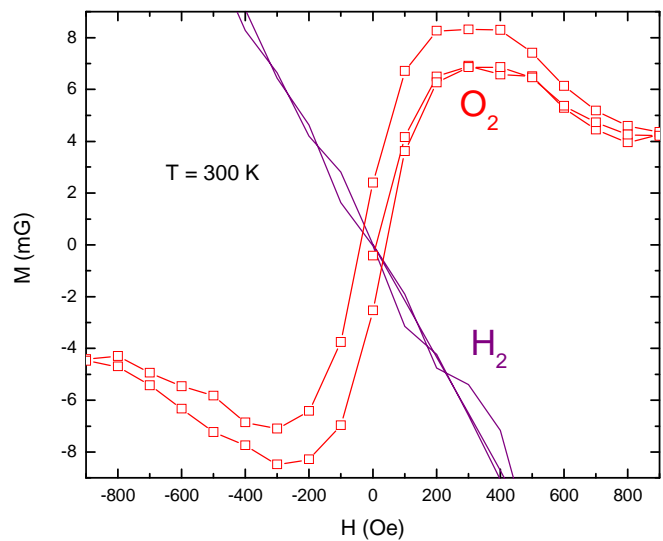

**Figure 5** (Color online) Magnetization M(H) measured at T = 300 K for the powdered graphite samples prepared in hydrogen ($H_2$) and oxygen ($O_2$) gas environments.

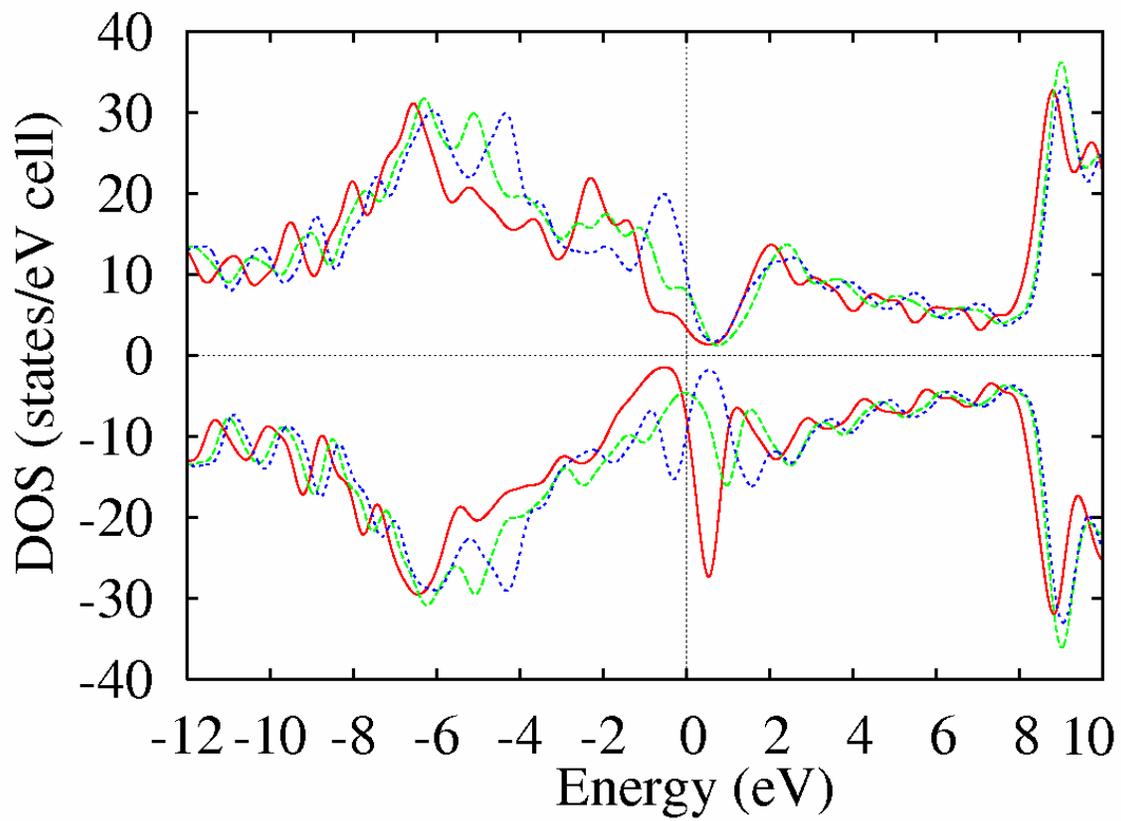

**Figure 6** (Color online) Spin-polarized densities of states for graphite with non-oxidized zigzag edges (solid red line) and with partially oxidized zigzag-edges. Dashed green line corresponds to the structure in Fig. 1e and the dotted blue line to the structure in Fig. 1f.